\documentclass[runningheads,fleqn]{llncs}

\usepackage{booktabs}   %% For formal tables:
\usepackage{subcaption} %% For complex figures with subfigures/subcaptions
\usepackage{latexsym}
\usepackage{amssymb}
\usepackage{amsmath}
\usepackage{stmaryrd}
\usepackage[all]{xy}
\usepackage{graphicx}
\usepackage{hyperref}
\usepackage{multicol}
\usepackage{bm}

\usepackage{tikz}
\usetikzlibrary{calc,arrows,shadows,positioning}

\newcommand{\ignore}[1]{}

\usepackage{xcolor}

\newcommand{\blue}[1]{{\color{blue} #1}}

\newcommand{\nil}{[\:]}

\def \tuple#1{\langle #1 \rangle}

\long\def\comment#1{}

\newcommand{\cauder}{\textsf{CauDEr}}

\newcommand{\spawn}{\mathsf{spawn}}
\newcommand{\spawninst}{\mathsf{spawn\_inst}}
\newcommand{\send}{\mathsf{send}}
\newcommand{\deliver}{\mathsf{deliver}}
\newcommand{\rec}{\mathsf{receive}}

\usepackage{algorithm}
\usepackage{algorithmic}

\newcommand{\pids}{\mathsf{Pids}}
\newcommand{\logtrace}{\mathsf{LT}}
\newcommand{\mailbox}{\mathsf{MBox}}
\newcommand{\fetch}{\mathit{fetch}}
\newcommand{\store}{\mathit{store}}
\newcommand{\newref}{\mathit{new\_unique\_ref}}
\newcommand{\updatepids}{\mathit{update\_pids}}
\newcommand{\processmsg}{\mathit{process\_msg}}
\newcommand{\processnewmsg}{\mathit{process\_new\_msg}}

\newcommand{\trydeliver}{\mathit{try\_deliver}}
\begin{document}

\title{A Program Instrumentation for Prefix-Based Tracing in
Message-Passing Concurrency%
\thanks{This work has been partially supported by EU (FEDER) and
    Spanish MCI/AEI under grant 
    PID2019-104735RB-C41, by the \emph{Generalitat Valenciana}
    under grant Prometeo/2019/098 (DeepTrust), and by
    French ANR project DCore ANR-18-CE25-0007.}
}
%\subtitle{(work in progress)}

\titlerunning{A Program Instrumentation for Prefix-Based Tracing}

\author{Juan Jos\'e Gonz\'alez-Abril  \and Germ\'an Vidal}

\authorrunning{J.J. Gonz\'alez-Abril  and G. Vidal}

\institute{
MiST, VRAIN, Universitat Polit\`ecnica de Val\`encia\\
\email{juagona6@vrain.upv.es,gvidal@dsic.upv.es}
}

\maketitle

\begin{abstract}
	The execution of concurrent programs generally involves some degree 
	of nondeterminism, mostly due to the relative speeds of the 
	concurrent processes. As a consequence, reproducibility is
	often challenging.
	This problem has been traditionally tackled by a combination of 
	\emph{tracing} and \emph{replay}. 
	In this paper, we introduce a program instrumentation for
	\emph{prefix-based tracing} 	that combines
	both tracing and replay. In the general case, the program is 
	instrumented with a \emph{partial} trace, so that the execution first
	follows the partial trace (replay) and, then, proceeds 
	nondeterministically, eventually producing a trace of the complete
	execution as a side effect. Observe that traditional tracing and
	replay are particular cases of our approach when an empty
	trace is provided (pure tracing) and when a full trace is provided
	(pure replay), respectively.
\end{abstract}

%\keywords{concurrency, logging, causal-consistent
%  debugging}

%%%%%%%%%%%%%%%%%%%%%%%%%%%%%%%%%%%%%%%%%%%%%
\section{Introduction} \label{intro}

Message-passing concurrency mainly follows the so-called
actor model. At runtime, concurrent processes can only 
interact through message sending and receiving (i.e., there is
no shared memory). In this paper, we further consider that
communication is asynchronous. Each process has a local
mailbox (a queue) and each sent message is eventually stored
in the mailbox of the target process. Moreover, we consider
that processes can be dynamically spawned at runtime.
The programming language Erlang \cite{erlang} mostly follows
this model.\footnote{In practice, some Erlang
built-in's involve shared-memory concurrency,
but we will not consider them in this work.}

In this context, computations are typically nondeterministic
because of the relative speeds of processes. Consider, for
instance, three processes, $\mathsf{p1}$, $\mathsf{p2}$, and
$\mathsf{p3}$. If $\mathsf{p1}$ and $\mathsf{p2}$ both 
send a message to process $\mathsf{p3}$, the order in which these
messages are received is not fixed. Here, we say that these
messages \emph{race}. Considering all alternatives for 
message races is the purpose of state-space exploration 
techniques like \emph{stateless model checking} \cite{God97} 
or \emph{reachability testing} \cite{LC06}. Intuitively speaking, 
these techniques start with an arbitrary execution, then consider 
some message race in this execution and, then, replay the
execution up to the message race but then consider a different
alternative. Executing a program deterministically up to a point
and then let it proceed nondeterministically is called 
\emph{prefix-based testing} in \cite{LC06}. Other approaches, 
like \cite{CGS13}, follow a similar process by inserting 
preemptive points in the code and, then, forcing the program
to follow a particular \emph{interleaving} up to a given point.
%then proceeding nondeterministically.

The nondeterminism of concurrent programs is also problematic
for program debugging, since reproducing 
a  buggy execution in a debugger is often challenging.
To overcome this problem, so-called \emph{record-and-replay} 
debuggers are often used. Here, the program is first instrumented 
in order to produce a \emph{trace} of the execution as a side effect.
Then, if some execution exhibits an incorrect behavior, the user
can load both the program and the trace into a replay debugger
in order to reproduce the considered computation.  This is the
case, e.g., of the \emph{reversible} debugger
\cauder\ for Erlang programs \cite{LNPV18,GV21}. 
%Similarly to other reversible debuggers,
%executions can be explored both forward and backward.
%However, i
In contrast to other debuggers, \cauder\ is
\emph{causal-consistent}, which means that an execution is
not reproduced in exactly the same order as the original,
recorded one; rather, it allows the user to focus on
the actions of a particular process, so that the actions of
other processes are only performed if there exists a
dependency with the actions of the considered process.
Replay debugging
in \cauder\ is driven by the \emph{trace} of an execution,
i.e., by the sequence of actions that must be performed
by each process. A trace can be seen as a \emph{partial order}
(i.e., it represents the class of all interleavings 
which give rise to causally equivalent computations).

In this work, we propose a new program instrumentation
that can be used both in the context of state-space exploration methods
and record-and-replay debugging. Our technique, called 
\emph{prefix-based tracing}, takes a program and a (possibly partial)
execution trace as input. When executing the instrumented 
program in the standard runtime environment, each process will
follow the considered trace until all the actions in this trace have 
been performed.
From this point on, the program will proceed nondeterministically,
eventually producing a trace of the complete execution as a 
side effect. 
Observe that traditional tracing and replay are particular cases 
of our approach when an empty trace is provided (pure tracing) 
and when a full trace is provided (pure replay), respectively.

%%%%%%%%%%%%%%%%%%%%%%%%%%%%%%%%%%%%%%%%%%%%%
\section{Prefix-Based Tracing: (Partial) Logs and Traces} \label{sec:tracing}

In this section, we introduce some notions on tracing for
message-passing concurrent programs.
Here, we consider an Erlang-like language with asynchronous
message passing. In the following, we focus on the concurrent
component of the language and omit the evaluation of
expressions (that follows an eager functional semantics;
see, e.g., \cite{LNPV18jlamp}).
We consider the following concurrent actions:
\begin{itemize}
\item $\spawn(\mathit{mod},\mathit{fun},\mathit{args})$, which
is used to dynamically create a new process to evaluate
function $\mathit{fun}$ (defined in module $\mathit{mod}$) 
with arguments $\mathit{args}$ (a list). E.g., 
$\spawn(\mathit{test},\mathit{client},[S,c1])$ spawns a  process 
that evaluates the expression $\mathit{client}(S,c1)$, where function
$\mathit{client}$ is defined in module $\mathit{test}$.\footnote{As 
in Erlang, functions and atoms (constants) 
begin with a lowercase letter while variables
start with an uppercase symbol. The language has no user-defined
data constructors, but allows the use of \emph{lists}---following the usual 
Haskell-like notation---and \emph{tuples} of the form
$\{e_1,\ldots,e_n\}$, $n\geq 1$ (a polyadic function).}
A $\spawn$ expression reduces to a fresh identifier, called pid
(for p\emph{rocess} id\emph{entifier}),
that uniquely identifies the new process.

\item $p \:!\: v$, which sends the value $v$ (the \emph{message}) 
to process $p$ (a pid). The expression reduces to $v$ and eventually
stores this value in the mailbox of process $p$ as a side effect.
Sending a message is an asynchronous operation, so the process
continues immediately with the evaluation of the next expression.

%\item $\mathsf{receive}~p_1~[\mathsf{when}~ g_1]  \to e_1; \ldots; 
%p_n~[\mathsf{when}~g_n]  \to e_n~ \mathsf{end}$, 
\item $\mathsf{receive}~p_1  \to e_1; \ldots; 
p_n  \to e_n~ \mathsf{end}$, which looks
for the \emph{oldest} message
in the process mailbox that matches some pattern $p_i$ 
and, then,
%the corresponding guard $g_i$ holds;\footnote{Guards are usually
%built from relational and arithmetic operators and are optional.}
%then, it 
continues with the evaluation of $e_i$.
When no message matches any pattern, execution is \emph{blocked}
until a matching message reaches
the mailbox of the process.
\end{itemize}
In order to model a running application, \cite{LNPV18jlamp}
introduces a \emph{global mailbox} $\Gamma$ that
represents the network (which is similar
to the notion of \emph{ether} in \cite{SFB10}). 
When a message is sent, it is first stored in $\Gamma$.
Typically, each message in $\Gamma$ is stored as a tuple,
including the pid of the sender,
the pid of the target process, and (in the tracing semantics)
a \emph{tagged} message. Tags were introduced in \cite{LPV19}
 in order to uniquely identify each message so that the sending
 and receiving of a message can be tracked.
The messages that are sent directly between two given processes are
stored in a single queue so that the order is kept. In contrast, 
when the senders are different, messages can 
reach the target process in any order. 

The semantics in \cite{LNPV18jlamp} has a rule for each 
concurrent action, together with an additional rule to
\emph{nondeterministically} deliver a message, i.e., move a message
from $\Gamma$ to the local (private) mailbox of the target
process. In general, given a process $\mathsf{p}$, we might have several 
queues of messages associated to different senders:
$\{\mathsf{p1},\mathsf{p},q_1\},\{\mathsf{p2},\mathsf{p},q_2\},
\ldots,\{\mathsf{p}n,\mathsf{p},q_n\}$, where $\mathsf{p}i$
are the pids of the sender processes and $q_i$ are the corresponding
queues of messages sent from $\mathsf{p}i$ to $\mathsf{p}$, 
$i=1,\ldots,n$.
This rule to deliver messages was removed from the semantics
of \cite{LPV19,LPV21} since the goal was to define a \emph{replay}
semantics and, in this case, the order in which the messages must be 
received is fixed by a given trace.

In the following, we consider that an execution \emph{trace} consists
of a collection of sequences of \emph{actions}, one per process.
The considered actions are the following:
\begin{itemize}
\item $\spawn(p)$, where $p$ is the pid of the spawned process;
\item $\send(\ell)$, where $\ell$ is the tag of the sent message, which is
(initially) stored in the global mailbox $\Gamma$;
\item $\deliver(\ell)$, where $\ell$ is the tag of the delivered message,
which is moved from $\Gamma$ to the local mailbox of the target process;
\item $\rec(\ell)$, where $\ell$ is the tag of the message consumed from
the local mailbox.
\end{itemize}
We note that $\deliver$ events are attributed to the target of the
message.

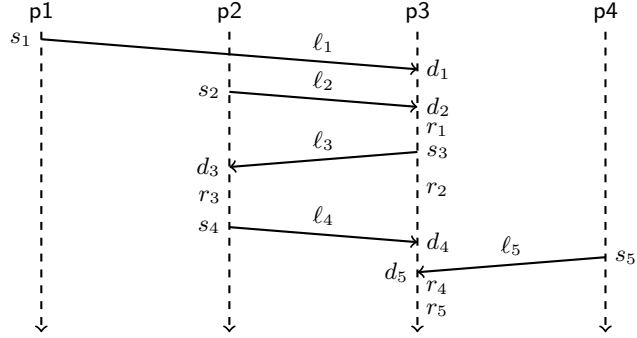
\begin{figure}[t]
\vspace{-5ex}
\centering
\begin{tikzpicture}
\draw[->,dashed,thick] (0,4) node[above]{\sf p1} -- (0,0);
\draw[->,dashed,thick] (2.5,4) node[above]{\sf p2} -- (2.5,0);
\draw[->,dashed,thick] (5,4) node[above]{\sf p3} -- (5,0);
\draw[->,dashed,thick] (7.5,4) node[above]{\sf p4} -- (7.5,0);
%\draw[->,dashed,thick] (10,4) node[above]{\sf p5} -- (10,0);

\draw[->, thick] (0,3.9) node[left]{$s_1$} -- (5,3.5) node[pos=0.75,above]{$\ell_1$} node[right]{$d_1$};

\draw[->, thick] (2.5,3.2) node[left]{$s_2$} -- (5,3) node[midway,above]{$\ell_2$} node[right]{$d_2$};

\node at (5,2.7) [right] {$r_1$};

\draw[->, thick] (5,2.4) node[right]{$s_3$} -- (2.5,2.2) node[midway,above]{$\ell_3$} node[left]{$d_3$};

\node at (5,1.9) [right] {$r_2$};

\node at (2.5,1.8) [left] {$r_3$};

\draw[->, thick] (2.5,1.4) node[left]{$s_4$} -- (5,1.2) node[midway,above]{$\ell_4$} node[right]{$d_4$};

\draw[->, thick] (7.5,1) node[right]{$s_5$} -- (5,0.8) node[midway,above]{$\ell_5$} node[left]{$d_5$};

\node at (5,0.6) [right] {$r_4$};

\node at (5,0.3) [right] {$r_5$};

\end{tikzpicture}
\caption{%Message-passing diagram. 
Processes ($\mathsf{p}i$, $i=1,\ldots,4$)  are represented
as vertical dashed arrows (time flows from top to bottom). 
Message sending and delivery is represented by solid arrows
labeled with a message tag ($\ell_i$),
from a sending event ($s_i$) to a delivery event ($d_i$), $i=1,\ldots,5$.
Receive events are denoted by $r_i$, $i=1,\ldots,5$. Note that all
events associated to a message $\ell_i$ have the same subscript $i$.}
\label{fig:simple}
\end{figure}

Let us consider the simple message-passing diagram
shown in Figure~\ref{fig:simple}. The associated trace can be 
represented as follows:
\[ \label{eqn:trace}
\begin{array}{ll}
\{\mathsf{p1},&[\blue{\spawn(\mathsf{p3})},
\blue{\spawn(\mathsf{p2})},
\blue{\spawn(\mathsf{p4})},\send(\ell_1)]\}\\
\{\mathsf{p2},&[\send(\ell_2),\deliver(\ell_3),\rec(\ell_3),\send(\ell_4)]\}\\
\{\mathsf{p3},&[\deliver(\ell_1),\deliver(\ell_2),\rec(\ell_1),\send(\ell_3),
\rec(\ell_2),\\
& ~\deliver(\ell_4),\deliver(\ell_5),\rec(\ell_4),\rec(\ell_5)]\}\\
\{\mathsf{p4},&[\send(\ell_5)]\}\\
\end{array}
\tag{1}
\]
Observe that we do not need to fix a particular interleaving
for all the actions in the trace. Only the order within
each process matters; i.e., a trace represents a partial order 
(analogously to the SYN-sequences of \cite{LC06}).

In \cite{LPV19,LPV21}, local mailboxes are abstracted away and 
there is no rule for message delivery
(messages are directly consumed from the global mailbox $\Gamma$).
Therefore, we have no $\deliver$ actions and the trace of the 
example above would be as follows:
\[ \label{eqn:log}
\begin{array}{ll}
\{\mathsf{p1},&[\blue{\spawn(\mathsf{p3})},
\blue{\spawn}(\mathsf{p2}),
\blue{\spawn(\mathsf{p4})},\send(\ell_1)]\}\\
\{\mathsf{p2},&[\send(\ell_2),\rec(\ell_3),\send(\ell_4)]\}\\
\{\mathsf{p3},&[\rec(\ell_1),\send(\ell_3),\rec(\ell_2),\rec(\ell_4),\rec(\ell_5)]\}\\
\{\mathsf{p4},&[\send(\ell_5)]\}\\
\end{array}
\tag{2}
\]
Despite the removal of $\deliver$ events, the resulting trace
(called \emph{log} in \cite{LPV19,LPV21}) suffices to replay
a given execution \cite[Theorem 4.22]{LPV21}. 
Nevertheless, the $\deliver$ events might be useful for
other purposes (e.g., to compute message races).

Therefore, in the following, we distinguish \emph{logs} 
(as in \cite{LPV19,LPV21}, without $\deliver$ events)
and \emph{traces} (with $\deliver$ events). For instance,
we say that the sequences in (\ref{eqn:trace}) represent a
trace while those in (\ref{eqn:log}) represent a log.

Observe that there are two message races in the execution
of Figure~\ref{fig:simple}. First, we have a race for $\mathsf{p3}$
between messages
$\ell_1$ and $\ell_2$. If we swap the delivery of these messages,
we can have a new execution which is not causally equivalent
to the previous one and, thus, may give rise to a different
outcome. A similar situation occurs with messages $\ell_4$ 
and $\ell_5$.
Here, one might be interested in considering a partial log in
order to explore an alternative execution. E.g., by assuming
that message $\ell_2$ reaches first process $\mathsf{p3}$,
we can produce the following partial log:
\[ \label{eqn:plog}
\begin{array}{ll}
\{\mathsf{p1},&[\blue{\spawn(\mathsf{p3})},
\blue{\spawn(\mathsf{p2})},
\blue{\spawn(\mathsf{p4})},\send(\ell_1)]\}\\
\{\mathsf{p2},&[\send(\ell_2)]\}\\
\{\mathsf{p3},&[\rec(\ell_2)]\}\\
\{\mathsf{p4},&[\send(\ell_5)]\}\\
\end{array}
\tag{3}
\]
Now, we could be interested in executing a program in replay
mode until the log is consumed and, then, continue nondeterministically,
eventually producing a trace of the complete execution. This is
the goal of \emph{prefix-based} tracing.

%The next section introduces a program instrumentation that
%achieves this goal.

%%%%%%%%%%%%%%%%%%%%%%%%%%%%%%%%%%%%%%%%%%%%%%%%%%%%%%%%%%
\section{The Program Instrumentation}\label{sec:inst} 

In this section, we focus on the design of a program instrumentation
to perform prefix-based tracing of (a subset of) Erlang programs.
In a nutshell, our program instrumentation proceeds
as follows:
\begin{itemize}
\item First, we introduce a new process, called the \emph{scheduler}
(a server), that will be run as part of the source
program.
\item The scheduler ensures that the actions of 
a given log are followed in the same order,
and that the corresponding trace is eventually computed.
It also includes a data structure that corresponds to the
global mailbox $\Gamma$. % introduced in the previous section.
In the instrumented program,
all messages will  be sent via the scheduler.
\item Finally, the sentences that correspond to the concurrent
actions $\spawn$, $\send$ and $\rec$ are instrumented
in order to interact with the scheduler. The remaining code will
stay untouched.
\end{itemize}
The scheduler uses several data structures called \emph{dictionaries}, 
a typical key-value data structure which is commonly used 
in Erlang applications. Here, we consider the following standard 
operations on dictionaries:
\begin{itemize}
\item $\fetch(k,dict)$, which returns the value $val$ 
associated  to key $k$ in $dict$. We write $dict[k]$ as a shorthand
for $\fetch(k,dict)$.

\item $\store(k,val,dict)$, which updates the dictionary
by adding (or updating, if the key exists) a new pair with 
key $k$ and value $val$. In this case, we write
$dict[k] := val$ as a shorthand for $\store(k,val,dict)$.
\end{itemize}
In particular, we consider the following dictionaries:
\begin{itemize}
\item $\pids$, which maps the pid of each process to a (unique) 
reference, i.e., $\pids[p]$ denotes the reference of pid $p$.
While pids are relative to a particular execution 
(i.e., the pid of the same process may change from one execution 
to the next one), the corresponding reference in a log or trace 
is permanent. 
This mapping is used to dynamically keep the association between
pids and references in each execution.
For instance, an example value for $\pids$ is
$[\{\tuple{0.80.0},\mathsf{p1}\}, 
\tuple{0.83.0},\mathsf{p2}\}]$, where $\tuple{0.80.0},\tuple{0.83.0}$ 
are Erlang pids and $\mathsf{p1},\mathsf{p2}$ are the corresponding
references.

\item $\logtrace$, which is used to associate each process reference
with a tuple of the form $\{ls,as\}$, where $ls$ is a (possibly empty)
list with the events of a log and $as$ is a (possibly empty) list
with the (reversed) trace of the execution so far.
The log is used to drive the next steps, while the second component
is  used to store the execution trace so far.
The list storing the trace is reversed for efficiency reasons
(since it is faster to add elements to the head of the list).
E.g., the initial value of $\logtrace$ for the partial
log displayed in (\ref{eqn:plog}) is as follows:
\[  
\begin{array}{ll}
[\{\mathsf{p1},&\{[\spawn(\mathsf{p3}),\spawn(\mathsf{p2}),
\spawn(\mathsf{p4}),\send(\ell_1)],\nil\}\},\\
~\{\mathsf{p2},&\{[\send(\ell_2)],\nil\}\},
\{\mathsf{p3},\{[\rec(\ell_2)],\nil\}\},
\{\mathsf{p4},\{[\send(\ell_5)],\nil\}\}]\\
\end{array}
\]

\item $\mailbox$, which represents the global
mailbox $\Gamma$. The key of this dictionary is the pid of
the target process, and the value is another dictionary in
which the keys are pids (those of the sender processes) and
the values are lists of (tagged) messages.
For instance, the value of $\mailbox$ after sending the first
two messages of the execution shown in Figure~\ref{fig:simple}
could be as follows:
%\footnote{For clarity, we use process references 
%here but the implementation considers pids.}
\[
\begin{array}{l@{~}l}
\{\tuple{0.84.0}, & \{\tuple{0.80.0},[\{\ell_1,v_1\}]\},\\
                          & \{\tuple{0.83.0},[\{\ell_2,v_2\}]\}\}
\end{array}
\]
where $\tuple{0.80.0},\tuple{0.83.0},\tuple{0.84.0}$
are the pids of $\mathsf{p1},\mathsf{p2},\mathsf{p3}$, respectively,
$v_1$ and $v_2$ are the message values and $\ell_1$ and $\ell_2$
are their respective tags.
\end{itemize}
Let us now describe the instrumentation of the source code.
First, every expression of the form 
$\spawn(\mathit{mod},\mathit{fun},\mathit{args})$ is replaced
by a call to a new function $\spawninst$ with the same arguments.
The new function is similar to the original function $\spawn$
but additionally (1) sends the message $\{P1,\mathsf{spawn},P2\}$
to the scheduler, where $P1$ is the pid of the current process
and $P2$ is the pid of the spawned process, and (2) inserts
a receive expression to make this communication \emph{synchronous}.
The reason for (2) is that
every message of the form $\{P1,\mathsf{spawn},P2\}$ must
add $P2$ to the data structure $\pids$, either with a new reference
or with the one in the current log. We require this operation to be
completed before either the spawned process or the one performing
the spawn can proceed with any other action. Otherwise, the
scheduler would run into an inconsistent state.

The instrumentation of message sending is much simpler. We just 
perform the following rewriting:
\[
e_1\:!\:e_2 \hspace{5ex} \Rightarrow \hspace{5ex}
\mathsf{sched}\:!\:\{\mathsf{self}(),\mathsf{send},e_1,e_2\}
\]
where $\mathsf{sched}$ is the pid of the scheduler
and $\mathsf{self}()$
is a predefined function that returns the pid of the current process.
%,
%$e_1$ is the expression that evaluates to the target pid, and $e_2$
%is the expression that evaluates to the message (a value).
%
Finally, the instrumentation of a receive expression rewrites the
code as follows:
\[
\begin{array}{l}
\mathsf{receive}~p_1  \to e_1; \ldots; 
p_n  \to e_n~ \mathsf{end}\\
\hspace{10ex}\Rightarrow 
\mathsf{receive}~\{L_1,p_1\}
\to \mathsf{sched}\:!\:\{\mathsf{self}(),\rec,L_1\},e_1; \ldots; \\
\hspace{20ex}\{L_n,p_n\}  
\to \mathsf{sched}\:!\:\{\mathsf{self}(),\rec,L_n\},e_n~ \mathsf{end}
\end{array}
\]
where $L_1,\ldots,L_n$ are fresh variables that are used to 
gather the tag of the received message and send it to the scheduler.

The main algorithm of the scheduler can be found in 
Algorithm~\ref{alg:scheduler}. First, we have an initialization
where the pid of the main process is associated with the
reference $\mathsf{p1}$ in $\pids$, the initial logs are 
assigned to $\logtrace$, and the mailbox is initially
empty. As is common in server processes, the scheduler is basically
an infinite loop with a receive statement to process the requests.
Here, we consider three requests, which correspond to the
messages sent from the instrumented source code.
Let us briefly explain the actions associated to each message:

\begin{algorithm}[t]
\caption{Scheduler} \label{alg:scheduler}
\begin{tabular}{l}
\textbf{Initialization}\\
\hspace{2ex} $\pids := [\{\mathsf{self}(),\mathsf{p1}\}]$;~
%\hspace{2ex} 
$\logtrace := 
   \mathtt{/*}~ \mathit{prefix}~\mathit{logs} ~\mathtt{*/}$; ~
%\hspace{2ex} 
$\mailbox := \{\:\}$;\\
\textbf{repeat}\\
~\textbf{receive}\\
\hspace{2ex} $\{p,\mathsf{spawn},p'\} \to$\\
%\hspace{8ex} $\updatepids([p,p'],\pids)$\\
%%\hspace{8ex} \textbf{case} $\fetch(\pids(p),\logtrace)$ \textbf{of}\\
\hspace{8ex} \textbf{case} $\logtrace[\pids[p]]$ \textbf{of}\\
\hspace{10ex} $\{\nil,as\} \to ~~
\mathtt{/*}~\mathit{trace~mode}~\mathtt{*/}$\\
\hspace{23ex} $r' := \newref()$;\\
\hspace{23ex} 
$\updatepids(p',r',\pids)$;\\
%%\hspace{23ex} $\store(\pids(p),\{\nil,[\spawn(r')|as]\},\logtrace)$;\\
\hspace{23ex} $\logtrace[\pids[p] :=\{\nil,[\spawn(r')|as]\}$;\\
\hspace{10ex} $\{[\mathsf{spawn}(r')|ls],as\} \to ~~
\mathtt{/*}~\mathit{replay~mode}~\mathtt{*/}$\\
\hspace{23ex} $\updatepids(p',r',\pids)$;\\
%%\hspace{23ex} $\store(\pids(p),\{ls,[\spawn(r')|as]\},\logtrace)$;\\
\hspace{23ex} $\logtrace[\pids[p]] := \{ls,[\spawn(r')|as]\}$;\\
\hspace{8ex} $p\: !\: ack$;\\
\hspace{8ex} 
$\trydeliver(p)$\\
\hspace{2ex} $\{p,\mathsf{send},p',v\} \to$\\
\hspace{8ex} \textbf{case} $\logtrace[\pids[p]]$ \textbf{of}\\
\hspace{10ex} $\{\nil,as\} \to ~~
\mathtt{/*}~\mathit{trace~mode}~\mathtt{*/}$\\
%\hspace{23ex} $\updatepids([p,p'],\pids)$\\
\hspace{23ex} $\ell := \newref()$;\\
\hspace{23ex} 
%%$\store(\pids(p),\{\nil,[\send(\ell)|as]\},\logtrace);$\\
$\logtrace[\pids[p]] := \{\nil,[\send(\ell)|as]\};$\\
\hspace{23ex} $\processnewmsg(\{p,p',\ell,v\},\mailbox,\logtrace);$\\
\hspace{10ex} $\{[\send(\ell)|ls],as\} \to ~~
\mathtt{/*}~\mathit{replay~mode}~\mathtt{*/}$\\
%\hspace{23ex} $\updateloggedpids([p,p'],[r,r'],\pids)$\\
%%\hspace{23ex} $\store(\pids(p),\{ls,[\send(\ell)|as]\},\logtrace);$\\
\hspace{23ex} $\logtrace[\pids[p]] :=\{ls,[\send(\ell)|as]\};$\\
\hspace{23ex} $\processmsg(\{p,p',\ell,v\},\mailbox,\logtrace);$\\
\hspace{8ex} $\trydeliver(p)$\\
\hspace{2ex} $\{p,\rec,\ell\} \to$\\
%%\hspace{8ex} \textbf{case} $\fetch(\pids(p),\logtrace)$ \textbf{of}\\
\hspace{8ex} \textbf{case} $\logtrace[\pids[p]]$ \textbf{of}\\
\hspace{10ex} $\{\nil,as\} \to ~~
\mathtt{/*}~\mathit{trace~mode}~\mathtt{*/}$\\
%\hspace{23ex} $\updatepids([p],\pids)$\\
%%\hspace{23ex} $\store(\pids(p),\{\nil,[\rec(\ell)|as]\},\logtrace)$\\
\hspace{23ex} $\logtrace[\pids[p]] := \{\nil,[\rec(\ell)|as]\}$\\
\hspace{10ex} $\{[\rec(\ell)|ls],as\} \to ~~
\mathtt{/*}~\mathit{replay~mode}~\mathtt{*/}$\\
%\hspace{23ex} $\updateloggedpids([p],[r],\pids)$\\
%%\hspace{23ex} $\store(\pids(p),\{ls,[\rec(\ell)|as]\},\logtrace)$\\
\hspace{23ex} $\logtrace[\pids[p]] := \{ls,[\rec(\ell)|as]\}$\\
\hspace{8ex} $\trydeliver(p)$\\
%~\textbf{end}\\
\textbf{until}~\textit{true}
\end{tabular}
\end{algorithm}

\begin{itemize}
\item If the message received has the form $\{p,\spawn,p'\}$,
where $p,p'$ are pids, we look for the tuple associated to process 
$\pids[p]$ in $\logtrace$. If the log is empty, we can proceed
nondeterministically and just need to keep a trace of the execution
step. Here, we obtain a fresh reference, $r'$, add the pair
$\{p',r'\}$ to $\pids$, and update the trace in $\logtrace$ with the new
action $\spawn(r')$. If the log is not empty, we proceed in a similar
way but the reference is given in the log entry.
Finally, we have to acknowledge the reception of this message since
this communication is synchronous (as explained above).
%In practice, we send back a message which includes a fresh constant
%to avoid confusion with other messages.

\item If the message received has the form $\{p,\send,p',v\}$,
we again distinguish the case where the process log is empty.
In this case, we obtain a fresh reference $\ell$ (the message tag) and
update $\logtrace$ with the new action $\send(\ell)$. 
Finally, we use the auxiliary function $\processnewmsg$ to 
check the log of the target process, $p'$, and then it proceeds as
follows:
\begin{itemize}
\item If the log of $\pids[p']$ is empty, we 
add the action $\deliver(\ell)$ to the
trace of $\pids[p']$ and then send the message to the target
process: $p'\:!\:\{\ell,v\}$, i.e.,  we apply an \emph{instant-delivery}
strategy, where messages are delivered as soon as possible
(this is the usual action in the Erlang runtime environment).
\item If the log is not empty, we do not know when this message
should be received. Hence, we add a new (tagged) 
message $\{\ell,v\}$ from $p$ to $p'$ 
to the mailbox $\mailbox$, and add an action
$\deliver(\ell)$ at the end of the current log. Note that 
computed logs (as in \cite{LPV19,LPV21}) should not
contain deliver actions. This one is artificially 
added to \emph{force} the delivery
of message $\ell$ as soon as possible (see function 
$\trydeliver$ below).
\end{itemize}
If the log is not empty, we proceed in a similar way but the message
tag is given by the log and we call the auxiliary function
$\processmsg$ instead. This function checks the log of the 
target process, $\pids[p']$, and then proceeds as follows: 
\begin{itemize}
\item If the next action in the log is $\rec(\ell)$, we add the action 
$\deliver(\ell)$ to the trace of $\pids[p']$ and  send the 
message to the target process: $p'\:!\:\{\ell,v\}$.
\item If the first action is not $\rec(\ell)$, we  add a new (tagged) 
message $\{\ell,v\}$ from $p$ to $p'$ to the mailbox 
$\mailbox$. Finally, if the log of $\pids[p']$ contains
an action $\rec(\ell)$, we are done; otherwise, an action
of the form $\deliver(\ell)$ is added to the end of the log
of process $\pids[p']$, as before.
\end{itemize}

\item Finally, when the received message has the form
$\{p,\rec,\ell\}$, we just update the trace with the new action
$\rec(\ell)$ and, if the log was not empty, we remove the first
action $\rec(\ell)$ from the log.
\end{itemize}
Each of the above cases ends with a call $\trydeliver(p)$, 
which is basically used to deliver messages that
could not be delivered before (because  it would have violated
the order of some log). For this purpose, this function  
checks the next action in the log of process $\pids[p]$.
If it has either the form $\rec(\ell)$ or $\deliver(\ell)$, and
the message tagged with $\ell$ is the oldest one
in one of the queues of $\mailbox$ with target $p$,
then we send the message to $p$, remove it from $\mailbox$
and add $\deliver(\ell)$ to the trace of process $\pids[p]$.
Furthermore, in case the element of the log was $\deliver(\ell)$,
we recursively call $\trydeliver(p)$ to see if there are more
messages that can be delivered.
In any other case, the function does nothing.

%%%%%%%%%%%%%%%%%%%%%%%%%%%%%%%%%%%%%%%%%%%%%%%%%%%%%%%%%%
\section{Concluding Remaks}\label{sec:concl} 

In this work, we have presented a program instrumentation for
prefix-based tracing of message-passing concurrent programs 
with asynchronous communication. An implementation to instrument
Erlang programs has been undertaken, which is publicly available
from\\
\url{https://github.com/mistupv/cauder/tree/paper/prefix-based-tracing}
.

As future work, we plan to first formalize a conservative extension
of the standard semantics in order to perform prefix-based
tracing (e.g., extending the logging semantics in \cite{LPV19,LPV21}).
Then, we will prove the correctness of the program instrumentation
w.r.t.\ the prefix-based tracing semantics. Finally, we plan to 
extend the causal-consistent reversible debugger \cauder\ in order
to use prefix-based tracing (e.g., as part of a
state-space exploration method).

\bibliographystyle{splncs04}
%\bibliography{biblio}

\end{document}